\documentclass[10pt, conference, letterpaper]{IEEEtran}

\usepackage{todonotes}
\usepackage{algorithm,algorithmic}
\usepackage{psfrag,pstool,textcomp,tabularx,subfigure}
\usepackage{amsmath}
\DeclareMathOperator*{\argmax}{arg\,max}
\DeclareMathOperator*{\argmin}{arg\,min}

\newcommand{\Dc}{\mathcal{D}}
\newcommand{\Lc}{\mathcal{L}}
\newcommand{\Pc}{\mathcal{P}}

\newcommand{\Zc}{\mathcal{Z}}
\newcommand{\Kc}{\mathcal{K}}
\newcommand{\Mcal}{\mathcal{M}}

\newcommand{\Fig}[1]{Fig.~\ref{fig:#1}}
\newcommand{\Sec}[1]{Sec.~\ref{sec:#1}}
\newcommand{\Tab}[1]{Tab.~\ref{tab:#1}}
\newcommand{\Eq}[1]{(\ref{eq:#1})}

\graphicspath{{figure/}{figures/}}

\begin{document}
\title{Multiservice UAVs for Emergency Tasks\\ in Post-disaster Scenarios}
%\titlenote{Produces the permission block, and copyright information}
%\subtitle{Extended Abstract}

\author{F. Malandrino$^\dagger$, C. Rottondi$^\ast$, C.-F. Chiasserini$^{\ast,\dagger}$, A. Bianco$^\ast$, I. Stavrakakis$^{\ddagger,\ast}$}
%\affiliation{%
 % \institution{$\dagger$ CNR-IEIIT, Italy $\quad$ $\ast$ Politecnico di Torino, Italy $\quad$ $\ddagger$ National and Kapodistrian Univ. of Athens, Greece}}
%\email{firstname.lastname@polito.it}

% The default list of authors is too long for headers}
%\renewcommand{\shortauthors}{F. Malandrino et al.}

\maketitle
\begin{abstract}
UAVs are increasingly being employed to carry out surveillance, parcel delivery, communication-support and other specific tasks.
Their equipment and mission plan are carefully selected to minimize the carried load an  overall resource consumption.
Typically, several single task UAVs are dispatched to perform different missions. In certain cases, (part of) the geographical area of operation may be common to these single task missions (such as those supporting post-disaster recovery) and it may
be more efficient to have multiple tasks carried out as part of a single UAV mission using common or even additional specialized equipment.

In this paper, we propose and investigate a joint planning of
multitask missions leveraging a fleet of UAVs equipped with a standard
set of accessories enabling heterogeneous tasks. To this end, an
optimization problem is formulated yielding the optimal joint planning
and deriving the resulting quality of the delivered tasks. In
addition, a heuristic solution is developed for large-scale
environments to cope with the increased complexity of the optimization
framework. The developed joint planning of multitask missions is
applied to a specific post-disaster recovery scenario of a flooding in
the San Francisco area. The results show the effectiveness of the proposed solutions and the potential savings in the number of UAVs needed to carry out all the tasks with the required level of quality.
\end{abstract}
 % The code below should be generated by the tool at
% http://dl.acm.org/ccs.cfm
% Please copy and paste the code instead of the example below.  %\begin{CCSXML}
%<ccs2012>
% <concept>
%  <concept_id>10010520.10010553.10010562</concept_id>
%  <concept_desc>Computer systems organization~Embedded systems</concept_desc>
%  <concept_significance>500</concept_significance>
% </concept>
% <concept>
%  <concept_id>10010520.10010575.10010755</concept_id>
%  <concept_desc>Computer systems organization~Redundancy</concept_desc>
%  <concept_significance>300</concept_significance>
% </concept>
% <concept>
%  <concept_id>10010520.10010553.10010554</concept_id>
%  <concept_desc>Computer systems organization~Robotics</concept_desc>
%  <concept_significance>100</concept_significance>
% </concept>
% <concept>
%  <concept_id>10003033.10003083.10003095</concept_id>
%  <concept_desc>Networks~Network reliability</concept_desc>
%  <concept_significance>100</concept_significance>
% </concept>
%</ccs2012>  
%\end{CCSXML} %\ccsdesc[500]{Computer systems organization~Embedded systems}
%\ccsdesc[300]{Computer systems organization~Redundancy}
%\ccsdesc{Computer systems organization~Robotics}
%\ccsdesc[100]{Networks~Network reliability}

% We no longer use \terms command
%\terms{Theory}

%\keywords{Unmanned Aerial Vehicles; Natural Disasters; Post-Emergency Monitoring; Fleet Area Coverage; Parcel delivery}

\section{Introduction}
The usage of Unmanned Aerial Vehicles (UAVs) to accomplish different
kinds of tasks in post-disaster recovery scenarios has recently become
the subject of investigation
\cite{erdelj2017wireless,erdelj2017help}. Fleets of UAVs performing
environmental monitoring \cite{kurz2011real,saeed2017realistic},
dispatching medicines in rural/hardly accessible areas
\cite{bamburry2015drones}, or ensuring mobile connectivity
\cite{fotouhi2018survey} have already been envisioned. As a relevant
example, UAVs are employed in Rwanda to deliver blood packs
to 21 hospitals located in remote and isolated areas on a regular
basis, even in the presence of harsh weather conditions \cite{ackerman2018medical}.

However, such critical tasks have up to now been considered in isolation, thus requiring separated fleets with equipment, computational resources, and capabilities dimensioned on the specific mission to be performed \cite{lee2017optimization}. 
In this study, we adopt a different approach and investigate a joint
planning of multitask missions leveraging a fleet of UAVs equipped
with a standard set of accessories (i.e., a videomonitoring system \cite{kurz2011real}, a cellular communication interface and a mounting frame for parcel carriage), which enables them to perform heterogeneous tasks (i.e., medicine/blood delivery, aerial monitoring, and mobile coverage). 
%As discussed above, the usage of a single fleet of multi-purpose UAVs instead that multiple fleets of specialized UAVs is expected to reduce the fleet size, while providing comparable service quality.

To show the benefits achieved by the usage of multi-purpose UAVs, we develop an optimization framework based on Integer Linear Programming (ILP) to optimally schedule their tasks in a post-disaster environment and apply it to a scenario of a simulated flooding event in the San Francisco area, where UAVs depart from one of the depots surrounding the emergency area and must return to a depot after completion of their task to change/recharge batteries. In addition,
a heuristic solution is developed for larger scale environments to
cope with the increased complexity of the optimization framework.

Results show that our
heuristic provides a performance closely approaching the
optimum. Furthermore, fully equipping all UAVs, e.g., providing all of
them with cameras and radios, allows for a greater flexibility that outweighs
the resulting lower payload available for parcel delivery missions, and further increases performance.

%The remainder of the paper is organized as follows. \Sec{related}
%briefly reviews the related literature. \Sec{model} presents an
%optimization formulation of the multitask UAVs problem. Then we present a heuristic approach to tackle large instances in \Sec{heuristic}, discuss our reference scenario in \Sec{scenario}, and report our numerical evaluation in \Sec{results}. Conclusions are drawn in the final section.

\section{Related Work}
\label{sec:related}

Beside military and security operations, the usage of UAVs is
envisioned in a plethora of civil applications, ranging from
agriculture to environmental monitoring and disaster management (see
\cite{otto2018optimization} for a thorough taxonomy and survey). In
the following, we focus on the three types of tasks encompassed in the
scenario under study.

\paragraph{UAV placement for wireless coverage}
UAVs can be leveraged in a number of wireless networking applications,
e.g., complementing existing cellular systems by providing additional capacity where needed, or to ensure network coverage in emergency or disaster scenarios (see \cite{mozaffari2018tutorial} for a comprehensive overview).
%In \cite{bupe2015relief}, an algorithm controlling the deployment and positioning of UAVs within cells of a mobile network has been implemented and demonstrated using quadcopters.
Differently from the works in \cite{mozaffari2018tutorial}, our model
%jointly considers a dynamic selection of the areas to be covered depending on the evolution of the disaster over time and based on the predicted mobility patterns of users. Additionally, it
jointly optimizes the scheduling of the UAV mobility and actions.
%moving, covering, monitoring, delivering and recharging actions.

\paragraph{UAV-based post-disaster monitoring systems}
As overviewed in \cite{chmaj2015distributed}, fleets of UAVs operating as distributed processing systems can be adopted for various monitoring tasks, including, e.g., surveillance, object detection, movement tracking, support to navigation. A prototype of UAV-based architecture for sensing operations has been described in \cite{yanmaz2018drone}. In our paper, we consider a conceptually similar UAV equipment of hardware and software modules.

\paragraph{UAVs for parcel delivery}
%Though the usage of UAVs for commercial delivery scopes has not yet become reality,
Several recent studies have already investigated optimization strategies for drone-assisted delivery models (see \cite{yoo2018drone} for a literature review). In particular, variations of the Travelling Salesman Problem leveraging UAVs for last-mile delivery have been introduced \cite{murray2015flying}.
%Focusing on medicine delivery, the authors of \cite{scott2017drone}
%compare two linear programming models that
%combine truck-based transportation and drone delivery.
%In our model, we consider a relatively small geographical area impacted by a natural disaster and focus on the last-mile drone assisted delivery problem, assuming that medicines have been carried beforehand at depots located at the border of the impacted area.

\begin{table}
\caption{Notation
    \label{tab:notation}
} %caption
\footnotesize
\begin{tabular}{|p{1.6cm}|p{1cm}|p{4.9cm}|}
\hline
Symbol & Type & Meaning \\
\hline\hline
$a(p)\in\Kc$ & parameter & Earliest epoch at which deliver payload~$p$\\
\hline
$b(p)\in\Kc$ & parameter & Latest epoch at which deliver payload~$p$\\
\hline
$C$ & parameter & Payload capacity of UAVs\\
\hline
$E$ & parameter & Battery capacity of UAVs\\
\hline
$e(l_1,l_2)$ & parameter & Energy consumed when traveling between locations~$l_1$ and~$l_2$, per unit of weight\\
\hline
$f(p)\in\Lc$ & parameter & Location at which payload~$p$ shall be delivered\\
\hline
$H$ & parameter & Horizon over which satisfaction is computed\\
\hline
$\Kc$ & set & Epochs\\
\hline
$\Lc$ & set & Locations\\
\hline
$\bar{\Lc}\subseteq\Lc$ & set & Locations with depots\\
\hline
$L(d,k)\in\Lc$ & Shorthand & Location of UAV~$d$ at epoch~$k$\\
\hline
$\Mcal$ & set & Non-delivery missions, e.g., coverage or monitoring\\
\hline
$n(k,m,z)$ & parameter & Work for mission~$m$ needed by users in zone~$z$ at epoch~$k$\\
\hline
$q(l,m,z)$ & parameter & Work for mission~$m$ that an UAV at location~$l$ can perform for users in zone~$z$, in one epoch~$k$\\
\hline
$r(m,p)\in\{0,1\}$ & parameter & Whether payload~$p$ is necessary to perform mission~$m$\\
\hline
$s(m)$ & parameter & Data generated by performing one unit of work of mission~$m$\\
\hline
$\Pc$ & set & Payload items\\
\hline
$\hat{\Pc}\subseteq\Pc$ & set & Payload items to be delivered\\
\hline
$t(l_1,l_2)$ & parameter & Traffic that can be transferred between locations~$l_1$ and~$l_2$, per epoch\\
\hline
$V$ & parameter & Maximum distance an UAV can cover in one epoch\\
\hline
$W$ & parameter & UAV weight\\
\hline
$w(p)$ & parameter & Weight of payload~$p$\\
\hline
$v(l_1,l_2)$ & parameter & Distance between locations~$l_1$ and~$l_2$\\
\hline
$\Zc$ & set & Zones\\
\hline
$\beta(d,k)$ & Real variable & Battery level of UAV~$d$ at epoch~$k$\\
\hline
$\lambda(d,k,l)$ & Binary variable & Whether UAV~$d$ is in location~$l$ at epoch~$k$\\
\hline
$\mu(d,k,m,z)\in[0,1]$ & Real variable & Fraction of epoch~$k$ that UAV~$d$ spends in mission~$m$ for zone~$z$\\
\hline
$\sigma(k,m,z)\in[0,1]$ & Real aux. variable & Satisfaction of users in zone~$z$ concerning mission~$m$ at epoch~$k$\\
\hline
$\bar{\sigma}(m)\in[0,1]$ & Real aux. variable & Mission-wide satisfaction concerning mission~$m$\\
\hline
$\tau(d_1,d_2,k)$ & Real variable & Traffic transferred from UAV~$d_1$ to UAV~$d_2$ at epoch~$k$\\
\hline
$\omega(d,k,p)$ & Binary variable & Whether UAV~$d$ carries payload~$p$ at epoch~$k$\\
\hline
\end{tabular}
\end{table}

\section{System model and optimization problem}
\label{sec:model}

%All indices are written between parentheses, in lexicographic order.

\paragraph{Space and time}
Time is discretized into a set~$\Kc=\{k\}$ of epochs, while space is discretized in a set~$\Lc=\{l\}$ of locationsThe notation\footnote{The notation we use is summarized
in \Tab{notation}.
Lower-case Greek letters indicate decision variables, lower-case Latin ones indicate parameters. Upper-case, calligraphic Latin letters indicate sets. Upper-case, regular Latin letters with indices indicate a specific element of the corresponding set, e.g., the location of an UAV. Upper-case, regular Latin letters without indices indicate design choices, e.g., UAV range, or system-wide parameters.}. The distance between two locations~$l_1$, $l_2$ is indicated as~$v(l_1,l_2)$ (clearly, $v(l,l)=0$). Some locations~$\bar{\Lc}\subseteq\Lc$ host depots.

Binary variables~$\lambda(d,k,l)$ indicate whether UAV~$d$ is at location~$l$ in epoch~$k$. Clearly, UAVs can only be in one location at a time and can only travel between locations closer than the maximum distance~$V$ UAVs can cover in an epoch. This translates into the following constraints:
\begin{equation}
\label{eq:one-location}
\sum_{l\in\Lc}\lambda(d,k,l)=1,\quad\forall d\in\Dc,k\in\Kc.
\end{equation}
\begin{equation}
\label{eq:close-travel}
\lambda(d,k,l)\leq\sum_{m\in\Lc\colon v(m,l)\leq V}\lambda(d,k-1,m)\quad\forall d\in\Dc,k\in\Kc,l\in\Lc.
\end{equation}

\paragraph{Payload}
UAVs have a payload capacity~$C$ and can carry zero or more payload
items~$p\in\Pc$, each weighting~$w(p)$. Examples of payload items (payloads for short) are blood packs or cameras. Binary decision variables~$\omega(d,k,p)$ express whether payload~$p$ is carried by UAV~$d$ at time~$k$.
\begin{equation}
\label{eq:capacity}
\sum_{p\in\Pc}w(p)\omega(d,k,p)\leq C,\quad\forall d\in\Dc,k\in\Kc.
\end{equation}
UAV payload can only change at depot locations:
\begin{equation}
\label{eq:no-payload-change}
\omega(d,k,p)=\omega(d,k-1,p),\quad \forall d\in\Dc,k\in\Kc,p\in\Kc\colon L(d,k)\notin\bar{\Lc}.
\end{equation}
\Eq{no-payload-change} implies that we do not account for the fact
that some payloads, e.g., medicine packs, will be dropped somewhere as
a part of the mission. This accounts for the worst-case event that one
or more drops fail, due to a variety of potential reasons (e.g.,
ground conditions are not adequate for UAV landing): in such a case, UAVs must have enough energy to bring all payloads back, if need be.

\paragraph{Energy and battery}
Real variables~$\beta(d,k)$ express the battery level of UAV~$d$ at epoch~$k$. Clearly, such variables shall be positive and can never exceed the battery capacity~$E$, i.e.,
\begin{equation}
\label{eq:b-range}
0\leq \beta(d,k)\leq E,\quad\forall d\in\Dc,k\in\Kc.
\end{equation}
Next, we need to account for power consumption:
\begin{multline}
\label{eq:b-consumption}
\beta(d,k)\leq \beta(d,k-1)+ \\
-e(L(d,k-1),L(d,k))\left(W+\sum_{p\in\Pc}\omega(d,k,p)w(p)\right),\\
\quad\forall d\in\Dc,k\in\Kc\colon L(d,k)\notin\bar{L}.
\end{multline}
In \Eq{b-consumption}, the energy consumed at time~$k$ is given by the product between a factor~$e(l_1,l_2)$, accounting for the distance between the locations, hence, for how far the UAV had to travel\footnote{Note that~$e(l,l)>0$, i.e., energy is also consumed by hovering over the same location.}, and the total weight of the UAV. Such a weight is given by the weight~$W$ of the UAV itself and the sum of the weight of the payload items it carries. Note that \Eq{b-consumption} does not hold at depot locations in~$\bar{\Lc}$, as UAVs can recharge or swap their batteries therein.

\paragraph{Delivery missions}
Some payload items~$\hat{\Pc}\subseteq\Pc$ must be delivered at certain location and times. Specifically, parameters~$f(p)\in\Lc$, $a(p)\in\Kc$, $b(p)\in\Kc$ indicate the target location (final point), as well as the earliest and latest times at which the delivery can take place. The following constraint imposes that all deliveries are carried out:
\begin{equation}
\label{eq:delivery}
\sum_{d\in\Dc}\sum_{k=a(p)}^{b(p)}\omega(d,k,p)\lambda(d,k,f(p))\geq 1,\quad\forall p\in\hat{\Pc}.
\end{equation}
\Eq{delivery} can be read as follows: there must be at least one epoch between~$a(p)$ and~$b(p)$ during which an UAV~$d$ visits the target location~$f(p)$ while carrying payload~$p$.

\paragraph{Additional missions}
We consider a set~$\Mcal=\{m\}$ of additional missions, e.g., wireless network coverage and monitoring. For the purposes of such missions, we partition the topology into zones~$z\in\Zc$, and express their demand for mission~$m$ at epoch~$k$ through parameters~$n(k,m,z)$, e.g., the traffic offered by the users\footnote{Notice that, for simplicity and without loss of generality, in this paper we only focus on uplink traffic.}. Parameters~$q(l,m,z)$ express how well an UAV in location~$l$ can perform mission~$m$ for zone~$z$, e.g., the quality of coverage it can provide. Furthermore, parameters~$r(m,p)\in\{0,1\}$ express the fact that some payload items~$p$, e.g., radios, are needed for mission~$m$. Finally, parameters~$s(m)$ express how much data is generated by performing one unit of work in mission~$m$.

The main decision to make is how long UAVs perform additional missions, and for the benefit of which zones. This is conveyed by variables~$\mu(d,k,m,z)\in[0,1]$, expressing the fraction of epoch~$k$ that UAV~$d$ uses to perform mission~$m$ for the benefit of zone~$z$. The first constraint we need to impose is that UAVs do not perform missions that they are not equipped for:
\begin{multline}
\label{eq:mission-equipment}
\mu(d,k,m,z)\leq \omega(d,k,p)\\
\quad\forall d\in\Dc,k\in\Kc,m\in\Mcal,p\in\Pc\colon r(m,p)=1,z\in\Zc.
\end{multline}
Also, we cannot exceed the need of zones:
\begin{multline}
\label{eq:mission-needs}
\sum_{d\in\Dc}\mu(d,k,m,z)q(L(d),m,z)\leq n(k,m,z)\\
\quad\forall k\in\Kc,m\in\Mcal,z\in\Zc.
\end{multline}
Note that \Eq{mission-needs} also accounts for the quality with which UAVs at different locations can perform the missions.

Next, we need to ensure that all the data traffic generated by
additional missions is transferred  to the in-field deployed cellular
network  (denoted with $\Omega$), so that it can be offloaded to the backbone network
infrastructure. We model such transfer to happen in a multi-hop
fashion, without store-carry-and-forward. We have a set of
parameters~$t(l_1,l_2)$ expressing the throughput that can be achieved
between UAVs staying at locations~$l_1$ and~$l_2$. If location~$l$ is
covered by a traditional network, then~$t(l,\Omega)$ expresses the amount of traffic that can be delivered to such a network in an epoch. Decision variables~$\tau(d_1,d_2,k)$ express the amount of data transferred from UAV~$d_1$ to UAV~$d_2$ at epoch~$k$.

We need to impose a  flow-like constraint, expressing that the
incoming traffic to every UAV~$d$, plus the one generated at~$d$ itself, must be transferred to either other UAVs or the traditional network:
\begin{multline}
\label{eq:flow}
\sum_{d'\in\Dc}\tau(d',d,k)+\sum_{m\in\Mcal}\sum_{z\in\Zc}\mu(d,k,m,z)q(L(d),m,z)s(m)=\\
=\sum_{d''\in\Dc}\tau(d,d'',k)+\tau(d,\Omega,k),\quad\forall d\in\Dc,k\in\Kc.
\end{multline}
We also need to account for the fact that only UAVs with specific equipment, e.g., a cellular radio, can act as relays. To this end, we add to the set of missions~$\Mcal$ an element called~$\textsf{relay}$, ensure that it requires the radio payload (i.e.,~$r(\textsf{relay},\textsf{radio})=1$), and then impose that only UAVs performing the~$\textsf{relay}$ mission act as relays:
\begin{multline}
\label{eq:max-trans}
\tau(d_1,d_2,k)\leq t(L(d_1),L(d_2))\mu(d_1,k,\textsf{relay},\cdot),\\
\quad\forall d_1\in\Dc,d_2\in\Dc\cup\{\Omega\},k\in\Kc.
\end{multline}
In \Eq{max-trans}, the $\cdot$~symbol {\em in lieu} of a zone indicates that the relay mission is specified for no particular mission. Also, \Eq{max-trans}  ensures that the maximum quantity of data that can be transferred~$t(L(d_1),L(d_2))$ is not exceeded.
Finally, all traffic generated by all missions must make its way to~$\Omega$:
\begin{multline}
\label{eq:all-omega}
\sum_{d\in\Dc}\sum_{m\in\Mcal}\sum_{z\in\Zc}\mu(d,k,m,z)q(l,m,z)s(m)=\sum_{d\in\Dc}\tau(d,\Omega,k),\\
\quad\forall k\in\Kc.
\end{multline}

\paragraph{Objective}
As a first step, we define the satisfaction~$\sigma(k,m,z)$ of zone~$z$ at epoch~$k$ for mission~$m$. Such a value is the ratio between how much service the zone was provided, and how much it needed. Importantly, it is not defined with reference to epoch~$k$ alone, but also to the previous~$H$ ones:
\begin{equation}
\label{eq:sigma}
\sigma(k,m,z)=\frac{\sum_{h=k-H}^k\sum_{d\in\Dc}\mu(d,k,m,z)q(L(d),m,z)}{\sum_{h=k-H}^k n(k,m,z)}.
\end{equation}

Leveraging the~$\sigma$ variables defined in \Eq{sigma}, we can define the mission-wise satisfaction as the minimum satisfaction across all zones and epochs:
\begin{equation}
\label{eq:sigma-bar}
\bar{\sigma}(m)=\min_{k\in\Kc}\min_{z\in\Zc}\sigma(k,m,z),\quad\forall m\in\Mcal.
\end{equation}

Finally, we can define our objective as maximizing the minimum satisfaction across all missions:
\begin{equation}
\label{eq:obj}
\max\min_{m\in\Mcal}\bar{\sigma}(m).
\end{equation}

\section{Heuristic Algorithm}
\label{sec:heuristic}

Focusing only on the blood/medicine delivery tasks, the problem
described in \Sec{model} can be modelled as a Vehicle Routing Problem
with Time Windows (VRPTW), which has been extensively studied in the
literature (see \cite{cordeau2000vrp} for a thorough overview on
heuristic and meta-heuristic approaches for VRPTW). 
In light of this, here  we present a heuristic algorithm aimed at tackling large instances of the considered post-disaster scenario, which builds upon the insertion method first proposed in \cite{solomon1987algorithms}. To incorporate additional tasks such as monitoring and connectivity coverage, we leverage the multi-objective enhancement of the insertion approach described in \cite{zografos2004heuristic}.

The insertion heuristic aims at sequentially building the tours of each UAV by adding one delivery location at a time. 
To do so, a graph is created where every delivery location $f(p) \in \Lc \colon p \in \hat{\Pc}$ is identified by a graph node $l$ (an additional node $l^{*}$ is added to identify the UAVs depot in $\bar{\Lc}$\footnote{For the sake of easiness, we assume that a single depot is used for all drones, i.e.,$|\bar{\Lc}|=1$ .}) and arc $(l,l')_{g}$ represent route $g\in G_{ll'}$ connecting delivery locations $l,l$. 
Note that we consider a set $G_{ll'}$ of alternative routes
for each location pair (i.e., every node pair is connected by $|G_{ll'}|$ arcs). 
More specifically, between each two locations~$l_1,l_2\in\Lc$, the following routes are considered:
\begin{itemize}
    \item the shortest path from~$l_1$ to~$l_2$;
    \item all paths going from~$l_1$ to an intermediate location~$l_3$
      and thence to~$l_2$, taking the shortest path between~$l_1$
      and~$l_3$ and the one between~$l_3$ and~$l_2$, provided that their length does not exceed twice that of the shortest path from~$l_1$ to~$l_2$;
    \item all paths including two intermediate locations~$l_3$
      and~$l_4$, subject to the same aforementioned conditions.
\end{itemize}

Each arc is associated with multiple weights: $\psi(l,l')_g$ and $e(l,l')_g$ respectively express the time (in number of epochs) and energy spent by the UAV to travel from node $l$ to $l'$ along route $g \in G_{ll'}$, whereas $c(l,l')_g$ and $\nu(l,l')_g$ respectively quantify the satisfaction level of coverage and monitoring tasks achieved by the UAV while travelling along route $g$ from $l$ to $l'$.
As tour initialization criterion, the insertion of the delivery task with earliest deadline has been chosen among the criteria proposed in \cite{solomon1987algorithms}. Then the algorithm iteratively operates as follows. Let $[l_0,l_1,...,l_m]$ be the current route, with $l_0,l_m=l^*$.
For each unserved delivery at $\overline{l} \in \Lc_u$ (where $\Lc_u \subseteq \Lc$ is the set of delivery locations not yet inserted in any tour), the best insertion position $\hat{i}_{\overline{l}} \in \{1,..,m\}$ is evaluated by minimizing the function $\phi_1(l_{i-1},\overline{l},l_{i})=\min_{g \in G_{l_{i-1},\overline{l}},g' \in G_{\overline{l},l_{i}} } (1-\alpha_1-\alpha_2) \cdot (\psi(l_{i-1},\overline{l})_{g}+\psi(\overline{l},l_{i})_{g'}-\psi(l_{i-1},l_{i})_{\hat{g}})-\alpha_1 \cdot (c(l_{i-1},\overline{l})_g+c(\overline{l},l_{i})_{g'}-c(l_{i-1},l_{i})_{\hat{g}})- \alpha_2 \cdot (\nu(l_{i-1},\overline{l})_g+\nu(\overline{l},l_{i})_{g'}-\nu(l_{i-1},l_{i})_{\hat{g}})$, where $\hat{g}$ is the route from $l_{i-1}$ to $l_{i}$ included in the current tour and $\alpha_1$,$\alpha_2$ are weights such that $\alpha_1+\alpha_2 \leq 1$. The closer $\alpha_1$ (resp. $\alpha_2$) approaches 1, the more predominant the satisfaction of coverage (resp. monitoring) tasks becomes w.r.t. the minimization of the total duration of the tour. Note that, if inserting the route pair $g,g'$ in the tour leads the overall energy consumption to exceed the UAV battery capacity or if the arrival epoch of the UAV at each delivery location does not meet the time window constraint of the corresponding delivery task, the insertion of the route pair $g,g'$ is considered as infeasible.

Once the value $\hat{i}_{\overline{l}}=\argmin_{i \in 1,..,m} \phi_1(l_{i-1},\overline{l},l_{i})$ has been found, in order to choose the best unserved delivery to be inserted in the tour, the function $\phi_2(l_{\hat{i}_{\overline{l}}-1},\overline{l},l_{\hat{i}_{\overline{l}}})=\max_{g \in G_{l^*,l_{\hat{i}_{\overline{l}}}}} (1-\alpha_1-\alpha_2)\cdot \psi(l^*,l_{\hat{i}_{\overline{l}}})-\alpha_1 \cdot c(l^*,l_{\hat{i}_{\overline{l}}})_g- \alpha_2  \cdot \nu(l^*,l_{\hat{i}_{\overline{l}}})_g-\phi_1(l_{\hat{i}_{\overline{l}}-1},\overline{l},l_{\hat{i}_{\overline{l}}})$ is computed for every unserved delivery $\overline{l} \in \Lc_u$. Such function quantifies the savings obtained by adding delivery $\overline{l}$ in the current tour, as opposed to direct service of delivery $\overline{l}$ in a new, dedicated tour starting from the depot. If $\max_{\overline{l} \in \Lc_u} \phi_2(l_{\hat{i}_{\overline{l}}-1},\overline{l},l_{\hat{i}_{\overline{l}}}) \geq 0$, then delivery $\hat{l}=\argmax_{\overline{l} \in \Lc_u} \phi_2(l_{\hat{i}_{\overline{l}}-1},\overline{l},l_{\hat{i}_{\overline{l}}})$ is added to the current tour and the insertion procedure is repeated from the start. Otherwise, a new tour is initialized. The algorithm ends when all the delivery tasks are inserted in a tour.

\section{Reference Scenario}\label{sec:scenario}

\begin{figure}
\centering
\includegraphics[width=1\columnwidth]{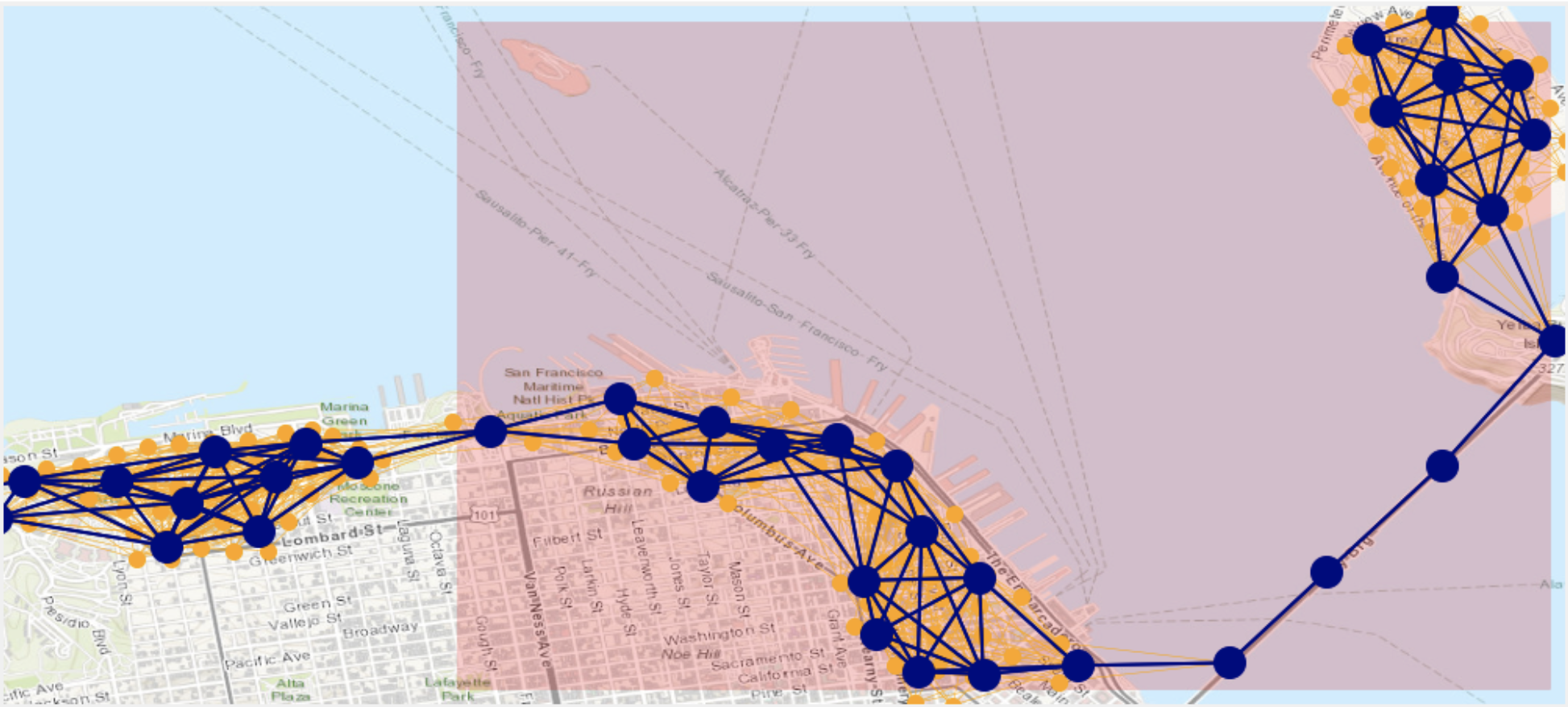}
\caption{
The reference topology we consider. Blue dots correspond to locations in~$\Lc$, while orange ones correspond to zones in~$\Zc$. Blue lines connect locations between which UAVs can travel in one epoch; orange lines connect zones with the locations from which UAVs can provide coverage to them. The shadowed area corresponds to the small-scale topology we use for our comparison against the optimum.
    \label{fig:topo}
} %caption
\end{figure}

As our reference scenario, we consider a flooding over San Francisco, depicted in \Fig{topo} and simulated through the software Hazus~\cite{hazus}. Over the disaster area, we identify~$|\Lc|=40$ locations and~$|\Zc|=50$ zones, with each zone reachable from an average of two locations. UAVs have to perform a total of~$|\hat{\Pc}|=20$ deliveries of blood or medicine packs, due at randomly-selected locations (the $f$-parameters) over a time window of 10~epochs for medicine packs and 5~epochs for blood packs (the~$a$- and $b$-parameters).

\begin{figure*}
\centering
\subfigure[\label{fig:onethird-perf}]{
    \includegraphics[width=.31\textwidth]{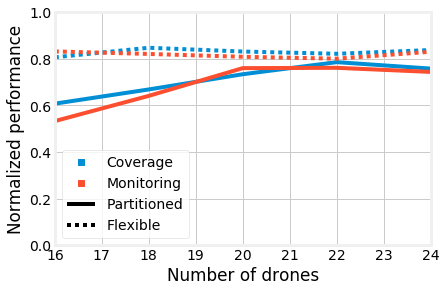}
} %subfigure
\subfigure[\label{fig:onethird-weight}]{
    \includegraphics[width=.31\textwidth]{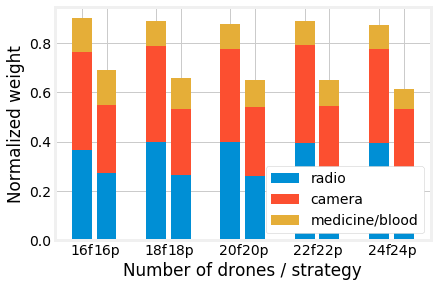}
} %subfigure
\subfigure[\label{fig:onethird-battery}]{
    \includegraphics[width=.31\textwidth]{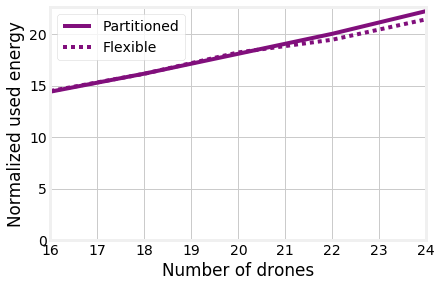}
} %subfigure
\caption{
    Small-scale scenario, optimal decisions: performance (a), payload (b), and used energy (c) yielded by flexible and fixed payload assignment strategies. Performance is normalized by the total demand, payload by the total capacity~$C$, and used energy by the battery capacity~$E$.
    \label{fig:onethird}
} %caption
\end{figure*}
\begin{figure*}
\subfigure[\label{fig:large-perf}]{
    \includegraphics[width=.32\textwidth]{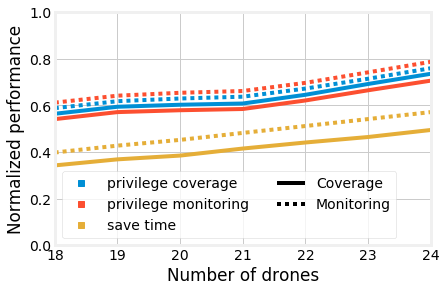}
} %subfigure
\subfigure[\label{fig:large-weight}]{
    \includegraphics[width=.32\textwidth]{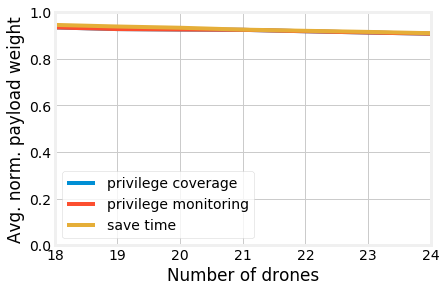}
} %subfigure
\subfigure[\label{fig:large-battery}]{
    \includegraphics[width=.32\textwidth]{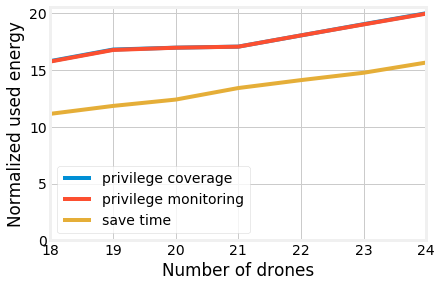}
} %subfigure
\caption{Large-scale scenario: performance (a), payload (b), and used energy (c) yielded by the heuristic strategy under different parameter settings. Performance is normalized by the total demand, payload by the total capacity~$C$, and used energy by the battery capacity~$E$.
    \label{fig:large}
} %caption
\end{figure*}
\begin{figure*}
\subfigure[\label{fig:small-perf}]{
    \includegraphics[width=.32\textwidth]{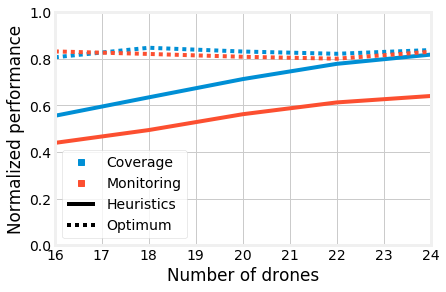}
} %subfigure
\subfigure[\label{fig:small-weight}]{
    \includegraphics[width=.32\textwidth]{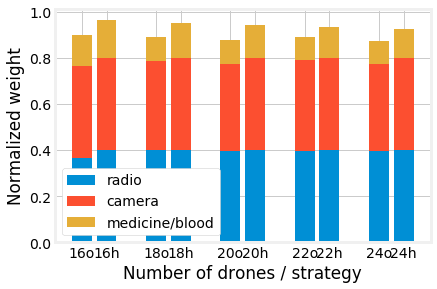}
} %subfigure
\subfigure[\label{fig:small-battery}]{
    \includegraphics[width=.32\textwidth]{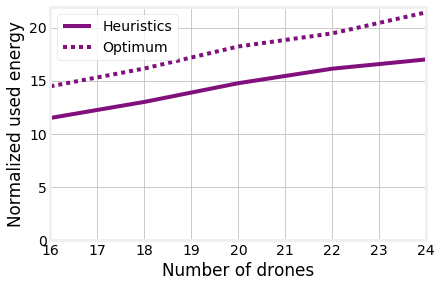}
} %subfigure
\caption{
    Small-scale scenario: performance (a), payload (b), and used energy, (c). Performance is normalized by the total demand, payload by the total capacity~$C$, and used energy by the battery capacity~$E$.
    \label{fig:small}
} %caption
\end{figure*}

UAVs can also perform $|\Mcal|=2$ additional missions:$i)$ providing network coverage for users escaping from the disaster, whose mobility is simulated through the MatSim simulator~\cite{matsim}, as detailed in~\cite{noi-workshop19}; $ii)$ video monitoring, e.g., to assess the level of the flooding in a certain area.

The quantity of needed service (the $n$-parameters) is determined as follows. For the coverage mission, the values computed in~\cite{noi-workshop19}, based on the expected flow of vehicles, are used. For video monitoring, a subset of 50 randomly-selected zones are deemed to need the service, hence, $n(z)=1$, while all others have~$n(z)=0$. Coverage and monitoring mission require additional payloads, respectively, the software radio~\cite{ettus} and the camera system~\cite{kurz2011real}, each weighting 1~kg. The maximum throughput values achievable between any two locations, i.e., the $t$-parameters, are obtained with reference to LTE micro-cells through the methodology in~\cite{noi-workshop19}; furthermore, it is assumed that UAVs can communicate with the ordinary network from all locations.

We consider a set of UAVs of variable cardinality, whose features mimic those of lightweight Amazon UAVs~\cite{xu2017design}. Specifically, they have an empty weight of~$W=4~\text{kg}$, and a maximum payload of~$C=2.5~\text{kg}$. They are equipped with a battery of capacity~$B=200~\text{Wh}$, and the energy consumed to fly between locations is~$e(l_1,l_2)=3.125~\text{Wh/km/kg}$. As a result, the range of an UAV carrying its maximum payload is around~$\frac{B}{e(C+W)}=9.8~\text{km}$. Interestingly, such a figure matches the 10-km range envisioned for lightweight UAVs in~\cite[Tab.~1]{stolaroff2018energy}.
Finally,  we consider~$|\Kc|=20$ epochs, each corresponding to 10~minutes, and a time horizon of~$H=10~\text{epochs}$.

\section{Numerical Assessment}\label{sec:results}

First, we seek to assess whether flexibility in the assignment of capabilities to drones, i.e., in deciding whether or not individual drones should carry a radio or a camera, translates into better performance. To this end, we consider the small-scale scenario represented by the shadowed area in \Fig{topo} and solve the problem presented in \Sec{model} to the optimum through an off-the-shelf solver. We consider two cases: (i) ``flexible'', where the equipment of drones is chosen by the optimizer, and (ii) ``fixed'', where additional constraints impose that one third of drones only carry the radio, one third only carry the camera, and one third carry both.

To this end, \Fig{onethird-perf} reports the fraction of the demand for coverage and monitoring that can be satisfied, as the number of available UAVs changes, and it clearly shows that flexibility results in substantially better performance. Interestingly, \Fig{onethird-weight} shows that, in the flexible case, drones are much more likely to carry both cameras and radios; indeed, recalling that cameras and radios weight 1~kg each and that the maximum payload is $C=$2.5~kg, we can conclude that drones do virtually always carry both. Additionally, as shown in \Fig{onethird-battery}, the global energy consumption is very similar in both cases: under the fixed strategy, the few drones equipped with cameras or radios are forced to take longer trips to provide a lower performance. This confirms our intuition that multiservice drones, equipped in a flexible manner, do indeed result in better performance.

Based on the results in \Fig{onethird}, we now configure our heuristic, described in \Sec{heuristic}, to always equip drones with both cameras and radios and assess its performance against the optimum.
In \Fig{large}, we consider the large-scale scenario depicted in \Fig{topo}, and study how the heuristic performs under different parameter settings. More in detail, we consider three settings: $\alpha_1=0, \alpha_2=0$ (\emph{save time}), $\alpha_1=1, \alpha_2=0$ (\emph{privilege coverage}), $\alpha_1=0, \alpha_2=1$ (\emph{privilege monitoring}). As reported in \Fig{large-perf}, privileging coverage or monitoring leads to similar overall performance in terms of service satisfaction, whereas the time saving substantially lowers the amount of offered coverage and monitoring. Different heuristic approaches have minor differences in terms of payload (\Fig{large-weight}), while the ``save time'' approach consumes substantially less energy than its counterparts, due to the shorter trip it results into.

Based on the above discussed results, wee now focus on the ``privilege coverage'' heuristic approach and compare its performance to the optimal ones, considering again the small-scale scenario depicted in \Fig{topo}.
As we can see from \Fig{small-perf}, the performance yielded by the heuristic is
remarkably close to the optimum, a significant fact given the
heuristic low complexity and high speed. It is also interesting to
observe how the difference between the coverage and monitoring
missions is smaller in the optimum than in the heuristic; indeed, when
decisions are made in a greedy fashion, it is harder to achieve a
perfect balance between coverage and monitoring. %Similarly, the fact
%that the heuristic occasionally provides more coverage than the
%optimum, does  {\em not} mean that its overall performance exceeds the
%optimum. Indeed, the objective function \Eq{obj} tends to maximize the minimum of the two dotted curves in \Fig{small-perf}.

\Fig{small-weight}, presenting the average weight of UAV payloads and
the breakdown thereof, provides an explanation for the performance
difference we can see in \Fig{small-perf}. Under the optimum strategy,
the payload carried by UAVs is always close to their capacity~$C$;
conversely, the heuristic tends to leave more free space. It follows
that UAVs can perform more deliveries in the same mission, visiting
more locations on the way. Interestingly, 
under the optimal strategy UAVs virtually always carry both cameras and radios, %This very important result confirms our intuition that having all UAVs equipped in the same way yields a better performance than having separate sets of special-purpose UAVs performing different missions. 
which validates our decision to equip all UAVs with both radio and camera in the heuristic approach.

\Fig{small-battery} shows how the total quantity of used energy, expressed in battery charges. Such a value is slightly higher under the optimal strategy than under the heuristic, a confirmation that heuristics trips tend to be shorter and visit fewer locations, thus performing fewer coverage and monitoring missions.

\section{Conclusions}
We addressed the challenging problem of jointly planning the missions
of multitask UAVs and has applied it to a post-disaster scenario. In
such cases, tasks are expected to be associated with a common
geographical area (i.e., the disaster area), hence UAVs carrying out
such tasks would largely geographically overlap. We show that
assigning multiple, instead of single, tasks to UAVs can lead to
savings in the number of UAVs required to carry out all the tasks,
provided that the problem of jointly planning the multiple tasks is
effectively addressed. To this end, we  developed an optimization
formulation and then an heuristic approach that effectively copes with
the computational complexity posed by the scenario. 
Using a realistic model of a flooding in the San
Francisco area and realistic
parameters for the operational 
equipment and tasks, we showed that our heuristic is a good match for
the  
optimum; moreover, the flexibility obtained by providing all UAVs with the same equipment translates into better performance.

\bibliographystyle{IEEEtran}
\bibliography{sigproc_free} 

\end{document}